\documentclass[preprint,12pt,nofootinbib]{revtex4}

\usepackage[brazil, english]{babel}
\usepackage{graphicx}
\usepackage{epsfig}
\usepackage{amssymb}
\usepackage{amsmath}
\usepackage{ulem}
\usepackage[T1]{fontenc} 
\usepackage[utf8]{inputenc}
\usepackage{color}
\usepackage{float}
\usepackage[titletoc]{appendix}
\usepackage[unicode=true,bookmarks=false,breaklinks=false,pdfborder={0 0 1},colorlinks=true]
 {hyperref}
\hypersetup{
 citecolor=blue,linkcolor=blue,urlcolor=blue}

\begin{document}

\title{Finite density effects on chiral symmetry breaking \\ in a magnetic field in 2+1 dimensions from holography}
 
\author{Diego M. Rodrigues$^{1,2} $}
\email[Eletronic address: ]{diegomhrod@gmail.com}
\author{Danning Li$^{3} $}
\email[Eletronic address: ]{lidanning@jnu.edu.cn} 
\author{Eduardo~Folco~Capossoli$^{4,5}$}
\email[Eletronic address: ]{eduardo\_capossoli@cp2.g12.br}
\author{Henrique Boschi-Filho$^{4}$}
\email[Eletronic address: ]{boschi@if.ufrj.br}  
\affiliation{$^1$Centro de Matemática, Universidade Federal do ABC, Santo Andr\'e, 09580-210, Brazil
\\$^2$ Centro de Física, Universidade Federal do ABC, Santo Andr\'e, 09580-210, Brazil
\\ $^3$Department of Physics and Siyuan Laboratory, Jinan University, Guangzhou 510632, China \\ $^4$Instituto de F\'isica, Universidade Federal do Rio de Janeiro, 21.941-972 - Rio de Janeiro-RJ - Brazil \\
$^5$Departamento de F\'\i sica and Mestrado Profissional em Pr\'aticas de Educa\c c\~ao B\'asica (MPPEB), Col\'egio Pedro II, 20.921-903 - Rio de Janeiro-RJ - Brazil}

\begin{abstract}
In this work we study finite density effects in  spontaneous chiral symmetry breaking as well as chiral phase transition under the influence of a background  magnetic field in $ 2+1 $ dimensions. For this purpose, we use an improved holographic softwall model based on an  interpolated dilaton profile. We find inverse magnetic catalysis at finite density. We observe that the chiral condensate decreases as the density increases, and the two effects (addition of magnetic field and chemical potential) sum up decreasing even more the chiral condensate. 
\end{abstract}


\maketitle

\newcommand{\limit}[3]
{\ensuremath{\lim_{#1 \rightarrow #2} #3}}

\section{Introduction}

Systems with finite chemical potential for fermions are a challenging and actual subject. The main reason for this interest is that in many physical systems we have to take into account a fermionic density, such as, in heavy-ion collisions, neutron stars, and condensed matter theories, among others. For a review, one can see, {\sl e. g.}, Refs. \cite{Rajagopal:2000wf, Boyanovsky:2001pt}.

In particular, nuclear matter can be treated within a very well-established theory, which is based on first principles and a nonperturbative approach, called lattice QCD (LQCD). The calculations in LQCD are numerical and usually via Monte Carlo simulations. However, at finite chemical potential LQCD seems to crash due to the sign problem, meaning that action of the theory becomes complex \cite{Goy:2016egl, Troyer:2004ge}.

Some proposals have been made to overcome this difficulty. For instance, in Refs.  \cite{Alford:1998sd, Giudice:2004se}, a complex chemical potential was used. In Refs. \cite{Barbour:1997ej, Fodor:2001au}, the authors have used reweighting approaches; in Ref. \cite{Blum:1995cb}, nonrelativistic expansions were used; and in Ref. \cite{Roberge:1986mm}, the authors have dealt with the reconstruction of the partition function. Very recently, in Ref. \cite{Borsanyi:2020fev},
the authors proposed a solution to this problem and provided extremely accurate results for the QCD transition, extrapolating from imaginary chemical potential up to real baryonic potential $\mu_{B}=300$ MeV.

There is an alternative approach to nonperturbative QCD, or even LQCD, based on the AdS/CFT correspondence \cite{Maldacena:1997re, Witten:1998zw}. Presented in 1997, this correspondence, generically referred to as holography, relates a strong coupling theory, without gravity, in a four-dimensional space to a weak coupling theory, including gravity, in a curved higher-dimensional space. The theoretical framework to deal with nuclear matter in the presence of a finite chemical potential within the AdS/CFT correspondence was put forward in many important works. See for instance,  Refs. \cite{Lee:2009bya, Jo:2009xr, Colangelo:2010pe, Colangelo:2011sr} and,  more recently, Refs. \cite{Bohra:2019ebj, Ghoroku:2020fkv, Evans:2020whc, Cao:2020ske, He:2020fdi, Zhou:2020ssi, Cao:2020ryx, Ballon-Bayona:2020xls, Mamani:2020pks, Braga:2019xwl, Braga:2020myi}.

Despite real QCD being a $(3+1)$-dimensional gauge theory the numerical calculations in a such a background (in the presence of a magnetic field) are extremely hard and reliable only for low values of the magnetic field, as can be seen in Refs. \cite{Dudal:2015wfn, Mamo:2015dea, Li:2016gfn, Evans:2016jzo, Chelabi:2015cwn, Chelabi:2015gpc, Bohra:2020qom} in the holographic context as well as in the nonholographic approach \cite{Bali:2011qj, Endrodi:2015oba}.

Here, in this work, our focus is to study the finite density effects on chiral symmetry breaking in the presence of a background constant magnetic field $B$ in $ 2+1 $ dimensions based on holographic studies done at zero density in Refs. 
\cite{Rodrigues:2017iqi,Rodrigues:2017cha,Rodrigues:2018pep, Rodrigues:2018chh}.
Our choice for a dimensional reduction comes from the fact that, in $ 2+1 $ dimensions, our model has a computational task easier than in real QCD even when  considering both nonzero chemical potentials and magnetic fields. This approach is very useful since we can learn from this model and try to extrapolate it to real QCD. Some previous nonholographic works in 2+1 dimensions with magnetic fields can be seen, for instance, in \cite{Klimenko:1990rh, Klimenko:1991he, Gusynin:1994re, Shovkovy:2012zn, Miransky:2015ava, Miransky:2002rp}.

This work is organized as follows. In section  \ref{holomodel} we describe our holographic model. In subsection \ref{geo} we detail the background  geometry which is an AdS$_4$-Reissner-Nordstrom black hole in the presence of a background magnetic field and present all relevant quantities for our further calculations. In the subsection \ref{effec} we show the holographic description of our effective action for a complex scalar field interacting with a non-Abelian gauge field. Such a complex scalar field will be related to the chiral condensate at the boundary theory. In section \ref{results} we present our numerical results where we observe inverse magnetic catalysis (IMC) at finite density as well as the decreasing of the chiral condensate when the density increases. In section \ref{conclusions} we make our conclusions.

\section{Holographic Model}\label{holomodel}

\subsection{Background geometry} \label{geo}
In this section we will establish the holographic description of our background geometry. Considering the Einstein-Maxwell action on $AdS_4$ (for more details see \cite{Rodrigues:2017cha,Rodrigues:2017iqi} and references therein):
\begin{equation} \label{AdS4Action}
S = \dfrac{1}{2\kappa^2_4}\int d^{4}x \sqrt{-g}\left(-\dfrac{6}{L^2} - L^{2}F_{MN}F^{MN}\right),
\end{equation}
where $ \kappa^2_4 $ is the four-dimensional coupling constant (with the relation $2\kappa^2_4\equiv16\pi G_4 $ with 4D Newton's constant $G_4$), $x^{M} = (t,z,x,y) $, with $ z $ being the holographic coordinate, and $F_{MN} = \partial_{M} A_N - \partial_{N} A_M$ is the field strength for the $U(1)$ gauge field $ A_{M} $. Throughout the text we will work in units such that $2\kappa^2_4 = L = 1$.

The field equations from Eq. \eqref{AdS4Action} are
\begin{eqnarray}
R_{MN} &=& 2\left(F_{M}^{P}F_{NP} - \dfrac{1}{4}g_{MN}F^2\right) - {3}g_{MN}, \label{FieldEquations}\\
\nabla_{M}F^{MN} &=& 0 \label{BianchiIdentity},
\end{eqnarray}
where $R_{MN}$ is the Ricci tensor and $g_{MN}$ is the metric tensor. Since we want to include a nonzero chemical potential $\mu$ and a constant magnetic field $B$, from now on in this work, it is worthwhile to mention that the magnetic field $B$ that we are considering is always a background field. 
The ansatz we are going to consider is the dyonic AdS/Reissner-Nordstrom black hole solution \cite{Hartnoll:2007ai,Hartnoll:2009sz}, with both electric and magnetic charge, given by:
\begin{eqnarray}
ds^2 &=& \dfrac{1}{z^2}\left(-f(z)dt^2 + \dfrac{dz^2}{f(z)} + dx^2 + dy^2\right), \label{AnsatzMetrica} \\
f(z) &=& 1-(1+\mu^2\,z_h^2+B^2\,z_h^4)\left(\dfrac{z}{z_h}\right)^3+(\mu^2\,z_h^2+B^2\,z_h^4)\left(\dfrac{z}{z_h}\right)^4 \label{horfunction},\\
A &=& \mu\,\left(1-\dfrac{z}{z_h}\right) \,dt + \dfrac{B}{2}(x\,dy-y\,dx). \label{ansatzA}
\end{eqnarray}

The temperature of the black hole solution can be obtained through the Hawking formula,
\begin{equation}\label{eq-T-1}
T = -\dfrac{f'(z_h)}{4\pi}, 
\end{equation}
and is given by
\begin{equation}\label{T}
T(z_h,\mu,B) = \dfrac{1}{4\pi\,z_{h}}\left(3 - B^2\,z_h^4 - \mu^2\,z_h^2\right). 
\end{equation}

Note that this equation for $z_h$ has four roots for fixed $T, \mu, B$, but only one is real and positive which means physically acceptable (note that by taking Eq.~\eqref{eq-T-1}, we have chosen the outer horizon). So, from now on, we just consider the branch $z_h >0$.

\subsubsection{IR near-horizon geometry: Emergence of AdS$_2$$\times\mathbb{R}^2$}

Here, we briefly discuss an important feature of the near-horizon ($z\to z_h$) geometry of extremal ($T=0$) charged black holes in asymptotically AdS spacetime, which is the emergence of the AdS$_2 \times\mathbb{R}^2$ space \cite{Edalati:2009bi,Faulkner:2009wj,Sachdev:2019bjn}. An extremal black hole is characterized by the fact that its temperature $ T $, Eq. \eqref{T}, is zero. In this case the horizon function, Eq.  \eqref{horfunction}, becomes
\begin{equation}
f(z)\Big|_{T=0} = 1-4\,\left(\frac{z}{z_h}\right)^3 + 3\,\left(\frac{z}{z_h}\right)^4,
\end{equation}
where it has a double zero at the horizon and can be Taylor expanded as
\begin{equation}
\setlength{\jot}{10pt}
\begin{aligned}
f(z)&\approx  \frac{6}{z_h^2}\,(z-z_h)^2.
\end{aligned}
\end{equation} 
Now, defining a dimensionless coordinate $ w $ through the rescaling $w := z/z_h$ and changing variables according to 
$ w = 1+z_h\,\eta$, we find that the geometry in the near-horizon region ($z\to z_h$) becomes
\begin{equation}
ds^2 \approx \left(- \frac{\eta^2}{L_{\mathrm{eff}}^2}\,dt^2 + \frac{L_{\mathrm{eff}}^2}{\eta^2}\,d\eta^2 \right) + \frac{1}{z_h^2}\,\left(dx^2+dy^2 \right),
\end{equation}
which is the AdS$_2 \times\mathbb{R}^2$, with AdS$_2$ curvature radius  $L_{\mathrm{eff}}\equiv 1/\sqrt{6}$, in units of $L=1$. Thus, in the near-horizon regime (IR), the AdS$_2 \times\mathbb{R}^2$ space controls the low-energy physics of the dual gauge theory on the boundary. 
Therefore, it seems that supergravity on AdS$_4$ flows in the IR to a gravity theory on AdS$_2$ which, in turn, is dual to a ($0+1$)-dimensional effective conformal quantum theory, which is referred to as an IR CFT$_1$. For a more extensive discussion on this topic, we refer the reader to Refs. \cite{Faulkner:2009wj,Sachdev:2019bjn}.

\subsection{Effective action for chiral symmetry breaking} \label{effec}

The effective action we consider to describe the chiral symmetry breaking is given by (we refer the reader to Refs. \cite{Rodrigues:2018chh, Rodrigues:2018pep} and references therein)
\begin{equation}\label{ChiralAction}
S = \dfrac{1}{2\kappa^2_4}\int d^{3}x \, dz \sqrt{-g}\,e^{-\Phi(z)}\mathrm{Tr}\left(D_{M}X^{\dagger}\,D^{M}X - V(X) - G^2 \right), 
\end{equation}
where $ X $ is a complex scalar field with mass squared $ M_4^2 = -2 $ dual to the chiral condensate $ \sigma\equiv\left\langle \bar{\psi}\psi\right\rangle $ in three spacetime dimensions, whose conformal dimension is $ \Delta=2 $. $ D_{M} $ is the covariant derivative defined as $ D_{M} \equiv \partial_{M} + i\mathcal{A}_{M} $, with $ \mathcal{A}_{M} $ being a non-Abelian gauge field, and its field strength $ G_{MN} $ is defined as $ G_{MN} \equiv \partial_{M}\mathcal{A}_{N} - \partial_{N}\mathcal{A}_{M} -  i[\mathcal{A}_{M},\mathcal{A}_{N}] $. $ V(X)$ is the potential for the complex scalar field $X$ given by $ V(X) = -2X^{2} + \lambda X^{4}$, where $ \lambda $ is the quartic coupling, which we will fix as $ \lambda = 1 $ from now on. This coupling allows the spontaneous and explicit chiral symmetry breakings to occur independently as pointed in Ref. \cite{Gherghetta:2009ac}. One should note that the magnetic field enters only as a background field on the AdS$_{4}$/RN geometry, and its contribution is encapsulated in the determinant of the spacetime metric $g$ in the effective action, and it does not couple directly to the complex scalar field nor the non-Abelian gauge field (there is no backreaction of the probe B-sensitive quark matter degrees of freedom).

Concerning the dilaton profile $ \Phi(z) $ appearing in Eq. \eqref{ChiralAction} we will consider \cite{Chelabi:2015cwn,Chelabi:2015gpc,Li:2016gfn}
\begin{equation}\label{dilatonprofilez}
\Phi(z) = - \phi_{0}z^2 + (\phi_{0} + \phi_{\infty})z^2\tanh(\phi_2\,z^2),
\end{equation}
having three parameters, which captures both IR and UV behaviors. It interpolates between the positive quadratic dilaton profile in the IR, $ \Phi(z\to\infty) = \phi_{\infty}\,z^2 $, $\phi_{\infty} > 0$, and the negative quadratic dilaton profile in the UV, $ \Phi(z\to0) = - \phi_{0}\,z^2 $, with $\phi_{0}>0$.
This dilaton field plays the role of a soft IR cutoff promoting the breaking of the conformal invariance.

Note that in Ref.\cite{Karch:2006pv} the authors proposed the soft wall model with a quadratic dilaton profile $\Phi(z)=a z^2$ with a positive constant $a$ reproducing the spectrum of vector mesons with linear Regge trajectories. In Ref. \cite{Karch:2010eg} the authors discuss the sign of the dilaton in soft wall models and claim that the constant $a$ should be positive; otherwise there will be unphysical massless vector mesons. We should emphasize that in our case there will be no unphysical mode since our dilaton in the IR regime is positive, i.e, $\phi(z) = k\,z^2$, with positive $k$.

On the other side, it was shown in Refs. \cite{Chelabi:2015cwn,Chelabi:2015gpc,Ballon-Bayona:2020qpq} for the positive sign of the quadratic dilaton $\Phi(z)=k z^2$, with positive $k$,  that  spontaneous chiral symmetry breaking cannot be reproduced. In our case, with the dilaton profile Eq.\eqref{dilatonprofilez}, we do not have this problem since it has a positive sign on the IR and a negative sign on the UV.

Assuming that the expectation value of the complex scalar field $X$ takes a diagonal form $ \left\langle X \right\rangle = \frac 12 \, \chi(z)\, I_2 $ for the $ SU(2) $ case \cite{Dudal:2015wfn,Li:2016gfn}, where $ I_2 $ is the $ 2\times2 $ identity matrix, the field equations for $\chi(z)$, derived from \eqref{ChiralAction}, are given by 
\begin{equation}\label{ChiralFieldEquations2}
\chi''(z) + \left(- \frac{2}{z} - \Phi'(z) + \frac{f'(z)}{f(z)}\right) \chi'(z) - \dfrac{1}{z^2 f(z)}\partial_{\chi}V(\chi) = 0,
\end{equation}
where $ ' $ means derivative with respect to $ z $, $f(z)$ is given by \eqref{horfunction}, and the potential becomes 
$ V(\chi)\equiv\mathrm{Tr}\,V(X) = -\chi^2 + \chi^4$.

The boundary conditions used to solve \eqref{ChiralFieldEquations2} are  \cite{Rodrigues:2018chh,Rodrigues:2018pep}: 
\begin{eqnarray}
	\chi(z) &=& m_{f}\,z +\sigma z^2+O(z^3)...,\quad z\to 0, \\
	\chi(z) &=& c_0 +\frac{c_0\,(4\,c_0^2 -2)}{z_h \left(B^2\, z_h^4+\mu^2\, z_h^2-3\right)}(z-z_h)+O((z-z_h)^2)...,\quad z\to z_h,
\end{eqnarray}
where $m_f$ is the source (fermion mass), and $\sigma$ is the chiral condensate. Moreover, $c_0$ is a coefficient obtained from evaluating Eq. \eqref{ChiralFieldEquations2} as a series expansion. Since we want to study spontaneous symmetry breaking, most of the results in this work will be derived with the source turned off, i.e., $m_f=0$. Then, in both the UV and IR sides, we have one undetermined coefficient, $\sigma$ and $c_0$ respectively. With certain values of the two coefficients, one can obtain the solutions $\chi_{\text{UV}}$ and $\chi_{\text{IR}}$ from both sides. Requiring $\chi_{\text{UV}}=\chi_{IR}$ and $\chi^\prime_{\text{UV}}=\chi^\prime_{\text{IR}}$, one obtains two equations and could solve out $\sigma$ and $c_0$.

In the next section, we will present our results concerning the chiral symmetry breaking at finite density and magnetic field as well as the phase diagram in the $\mu-T$ plane. For convenience, we will express the dilaton parameters in units of the mass scale $\sqrt{\phi_{\infty}}$ as well as all our results. To be more clear, note that one can define a dimensionless variable by rescaling the $z$ coordinate as $u:=\sqrt{\phi_{\infty}}\,z$, so that the dilaton profile \eqref{dilatonprofilez} takes the form
\begin{equation}
\Phi(u) = -\tilde{\phi_{0}}\,u^2+(1+\tilde{\phi_{0}})\,u^2\,\tanh(\tilde{\phi_2}\,u^2),
\end{equation}
where $\tilde{\phi_{0}}:=\frac{\phi_{0}}{\phi_{\infty}}$ and $\tilde{\phi_{2}}:=\frac{\phi_{2}}{\phi_{\infty}}$ are the dimensionless parameters. 

Furthermore, one can check that the tachyon equation \eqref{ChiralFieldEquations2} can also be put in the dimensionless form by redefining all the dimensional quantities, for instance, $(z_h,\mu, B)$, in units of $\phi_{\infty}$. In this way, we have a two-parameter dilaton profile controlled by the dimensionless parameters $(\tilde{\phi_{0}},\tilde{\phi_{2}})$. Also, note that $ \tilde{\phi_{0}} $ is just the ratio between the dilaton parameter in the UV, $\phi_0$, and the dilaton parameter in the IR, $\phi_{\infty}$. Finally, for reference in the next section, we fix our parameters as $\tilde{\phi_{0}}=4.675$ and $\tilde{\phi_{2}} = 0.0375$.

\section{Results}
\label{results}

In this section, we present our results concerning the chiral phase transition in $2+1$ dimensions at finite temperature and density, in the presence of a background magnetic field. 
It is important to note that, in our model, we just consider the deconfined phase of the dual gauge theory, since we are working in the finite temperature ansatz for the black hole. In a more realistic model, one should also consider a confined phase (thermal AdS) which appears for low temperatures. These two phases are separated by a Hawking-Page phase transition. However, we were able to extrapolate some of our results for very low temperatures within the deconfined phase in order to show that our model realizes spontaneous breaking of chiral symmetry.

All the physical quantities presented in this section have a tilde, meaning that they are in units of the mass scale $\sqrt{\phi_{\infty}}$. To be more precise:

\begin{equation}
	\setlength{\jot}{12pt}
	\begin{aligned}
		(\tilde T,\tilde\mu) \equiv \left(\frac{T}{\sqrt{\phi_{\infty}}},\frac{\mu}{\sqrt{\phi_{\infty}}}\right)\,\,\, {\rm and} \,\,\,(\tilde B, \tilde \sigma) \equiv \left(\frac{B}{\phi_{\infty}},\frac{\sigma}{\phi_{\infty}}\right).
	\end{aligned}
\end{equation}   

\begin{figure}[H]
	\centering
	\includegraphics[scale = 0.35]{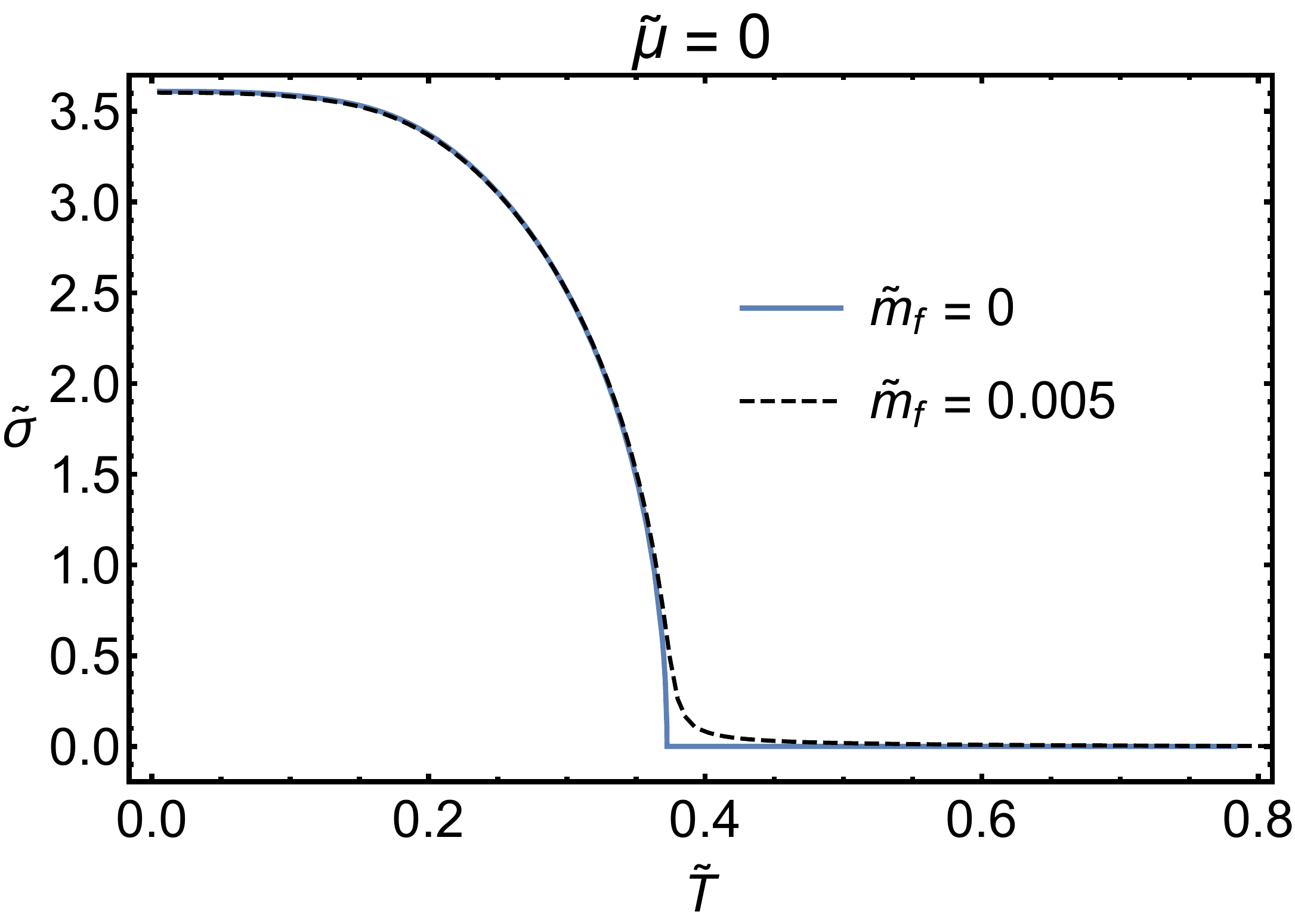}
	\hfill
	\includegraphics[scale = 0.35]{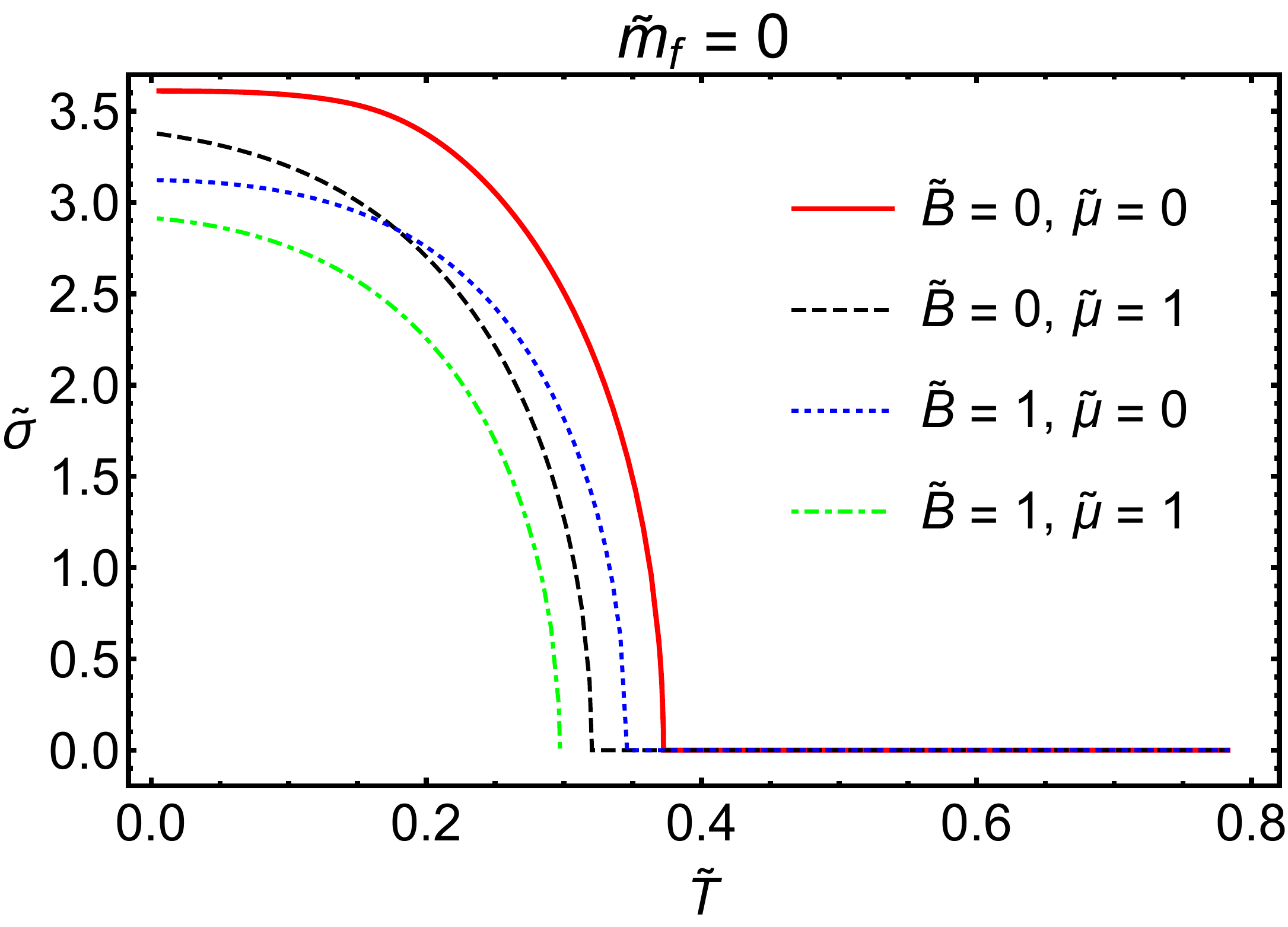}
	\caption{The chiral condensate $\tilde\sigma$ versus the temperature $\tilde T$. {\sl Left panel}: $\tilde\sigma$ versus $\tilde T$ for $\tilde\mu=0$ and $\tilde B=0$. In the chiral limit, i.e, $\tilde m_f=0$, one sees a second-order phase transition, while for $\tilde m_f\not=0$ there is a crossover. {\sl Right panel}: $\tilde\sigma$ versus $\tilde T$ in the chiral limit $\tilde m_f=0$ for different values of magnetic field and chemical potential. All quantities are in units of $\sqrt{\phi_{\infty}}$, in both panels.}
	\label{fig:svsTa}
\end{figure}

In Figure \ref{fig:svsTa} the behavior of the chiral condensate $\tilde{\sigma}$ as a function of temperature $\tilde{T}$ is presented. In the {\sl Left panel}, for zero magnetic field and density, one can see that the chiral phase transition is second-order in the chiral limit ($\tilde m_f=0$), while for finite fermion mass ($\tilde m_f\neq0$) the phase transition turns to a crossover. We have checked numerically that there is always spontaneous chiral symmetry breaking whenever the parameter $\tilde{\phi_{0}}\neq 0$. However, in the limit $\tilde{\phi_{0}}\to 0$ the chiral condensate vanishes. This limit corresponds exactly to the situation where the positive quadratic dilaton profile ($\phi(z) \sim \phi_{\infty}\,z^2$) dominates, and in this case it is known that the chiral symmetry breaking cannot be reproduced \cite{Chelabi:2015cwn,Chelabi:2015gpc,Ballon-Bayona:2020qpq}.

In the {\sl Right panel} of Figure \ref{fig:svsTa}, the chiral condensate $\tilde{\sigma}$ as a function of the temperature $\tilde{T}$ for different values of magnetic field and density is presented. At zero magnetic field and finite density, the value of the chiral condensate is reduced from its value at both zero magnetic field and density. At zero density and finite magnetic field, this reduction is observed, in agreement with previous works \cite{Rodrigues:2018chh,Rodrigues:2018pep}, signaling an inverse magnetic catalysis (IMC) effect. 
At finite density, with or without magnetic field, we observe a reduction of the condensate with respect to the condensate at zero density and magnetic field. This is expected to happen since the introduction of a chemical potential generates an asymmetry \cite{Preis:2010cq, Preis:2012fh} between the fermions ($\Psi$) and antifermions ($\bar{\Psi}$) which renders difficult the pairing $\bar{\Psi}\Psi$, i.e., the formation of a chiral condensate. At finite density and magnetic field there is also an IMC due to a summation of the effects such that the net effect is a reduction of the chiral condensate.\footnote{This reduction in the chiral condensate at finite density also appears in more sophisticated holographic approaches in higher dimensions, see for instance \cite{Gursoy:2017wzz,Ballon-Bayona:2017dvv}. Furthermore, for an alternative interpretation of IMC at vanishing chemical potential, based on the anisotropy caused by a magnetic field, see \cite{Gursoy:2018ydr}.}

\begin{figure}[H]
	\centering
	\includegraphics[scale = 0.35]{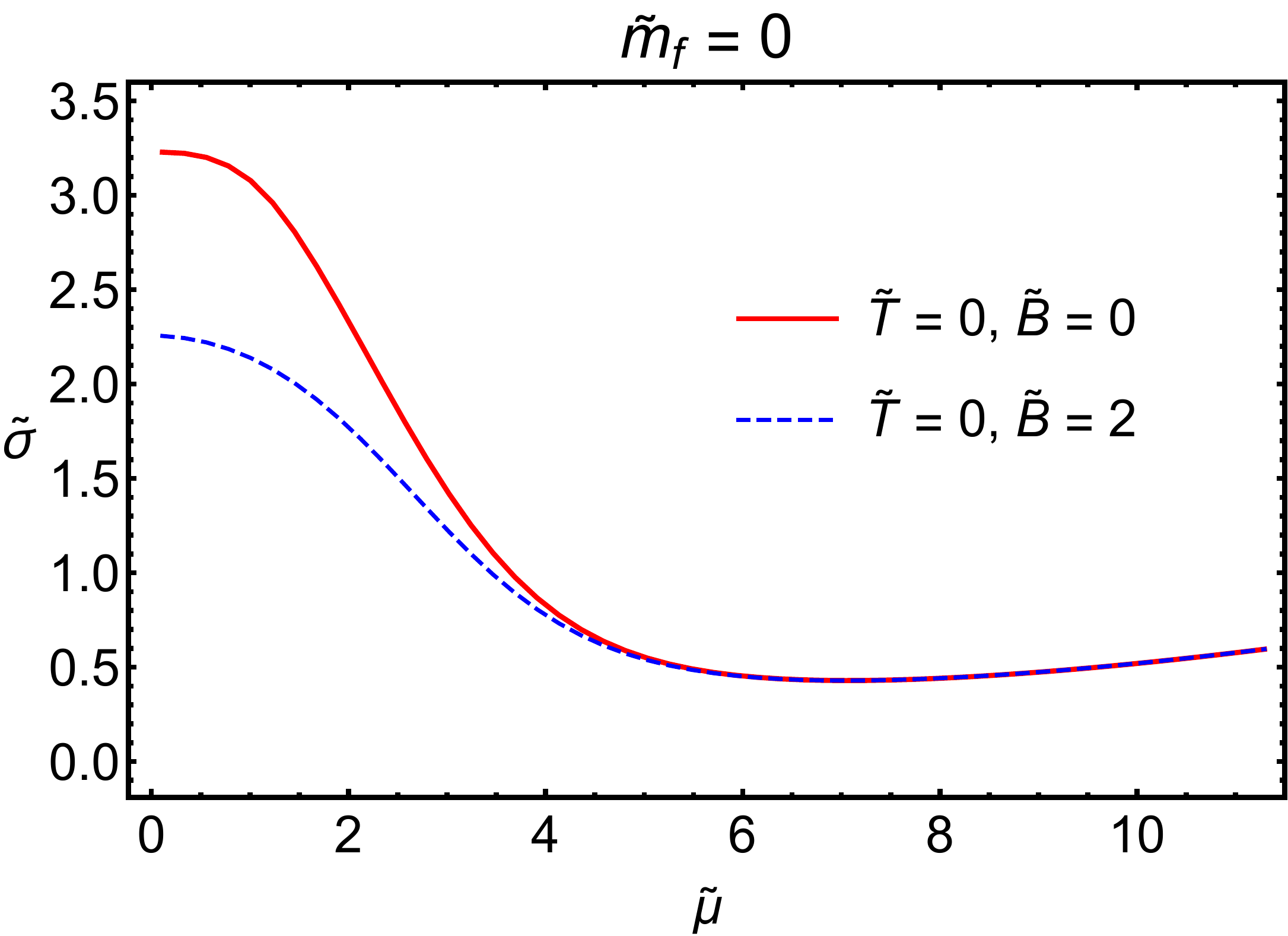}
	\hfill
	\includegraphics[scale = 0.35]{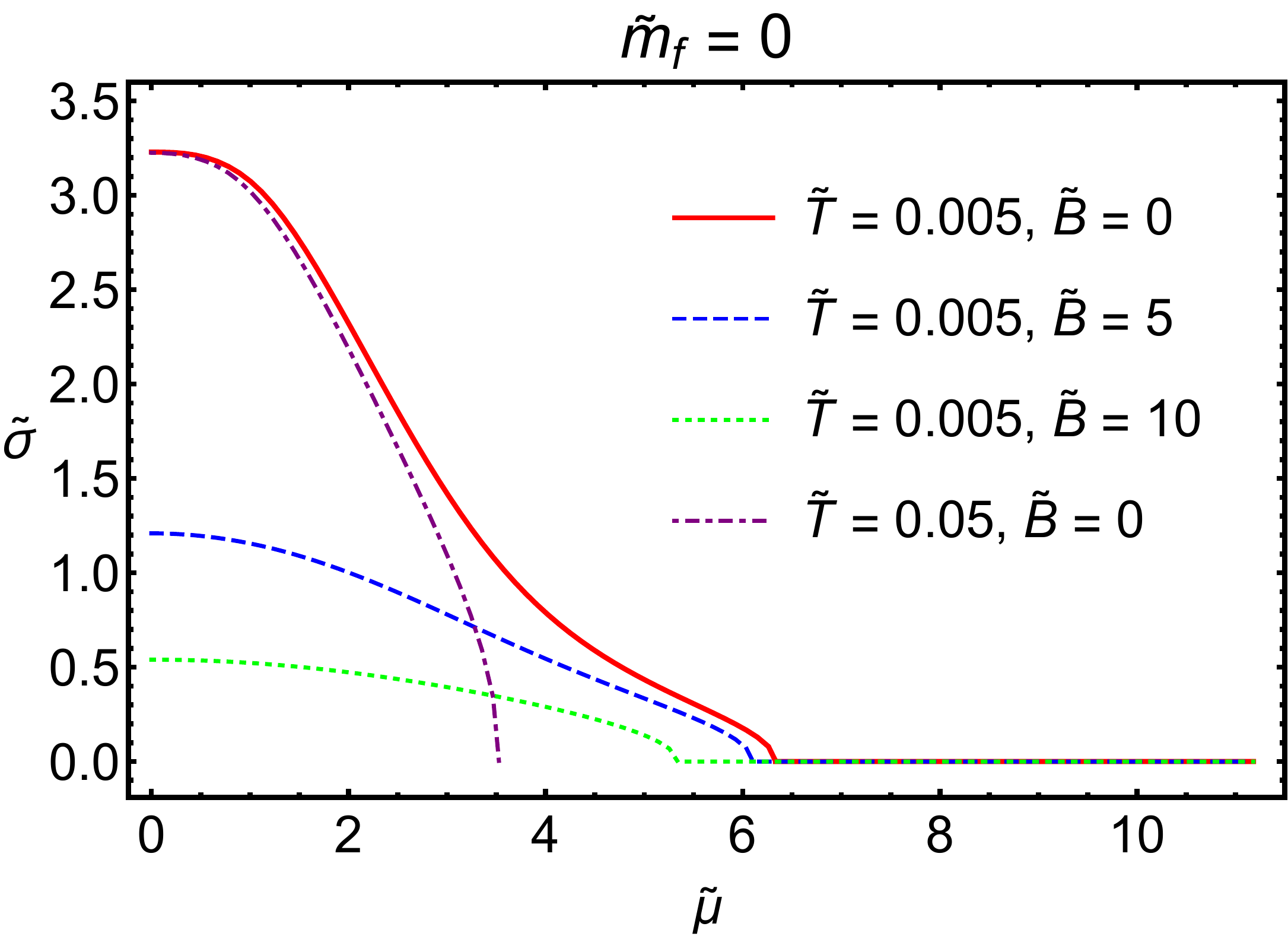}
	\caption{The chiral condensate $\tilde \sigma$ versus the chemical potential $\tilde\mu$ in the chiral limit $\tilde m_f=0$. {\sl Left panel}: $\tilde\sigma$ versus $\tilde\mu$  at zero temperature and different values of magnetic field. {\sl Right panel}: $\tilde\sigma$ versus $\tilde \mu$ at small finite temperatures and different values of magnetic field. All quantities are in units of $\sqrt{\phi_{\infty}}$, in both panels.}
	\label{fig:svsTb}
\end{figure}

In Figure \ref{fig:svsTb}, the chiral condensate as a function of the chemical potential in the chiral limit $\tilde m_f=0$ is shown. In the {\sl Left panel} this behavior is shown at zero temperature. One can see that the finite density affects the chiral condensate destructively, i.e., causing a decreasing,  until a critical chemical potential from which the chiral condensate starts to increase slowly again at large chemical potential.

As two comments, in the context of canonical (nonholographic) QCD, it is worthy to point out first that at large chemical potentials the chiral symmetry is expected to be restored. Second, at extremely high densities ($\mu>>T$) chiral symmetry can be broken through the formation of a condensate of quark Cooper pairs in the color-flavor-locked (CFL) phase, via a different mechanism \cite{Alford:1998mk, Alford:2007xm}. Note that this very mechanism was  discussed within the AdS/CFT program, for instance, in Ref.  \cite{Chen:2009kx}.

In the {\sl Right panel} of Figure \ref{fig:svsTb}, the chiral condensate as function of the chemical potential in the chiral limit $\tilde m_f=0$ for zero and finite magnetic fields is shown at finite small temperatures. One can observe in this case that the thermal effects affect the chiral condensate substantially, which is expected since thermal fluctuations have a huge impact on the chiral condensation, especially in $2+1$ dimensions \cite{Das:1995bn}. Moreover, with a finite magnetic field turned on, the decrease of the chiral condensate is much more pronounced, even at low temperatures, characterizing an IMC. Note, however that, at low temperatures, IMC is not expected to happen in QCD, since it is known that a magnetic field is a strong catalyst of chiral symmetry breaking, and, therefore, magnetic catalysis is expected to dominate in this low-temperature regime and, in particular, is universal behavior at zero temperature \cite{Gusynin:1994re,Miransky:2002rp,Shovkovy:2012zn,Miransky:2015ava}.

\begin{figure}[H]
	\centering
	\includegraphics[scale = 0.35]{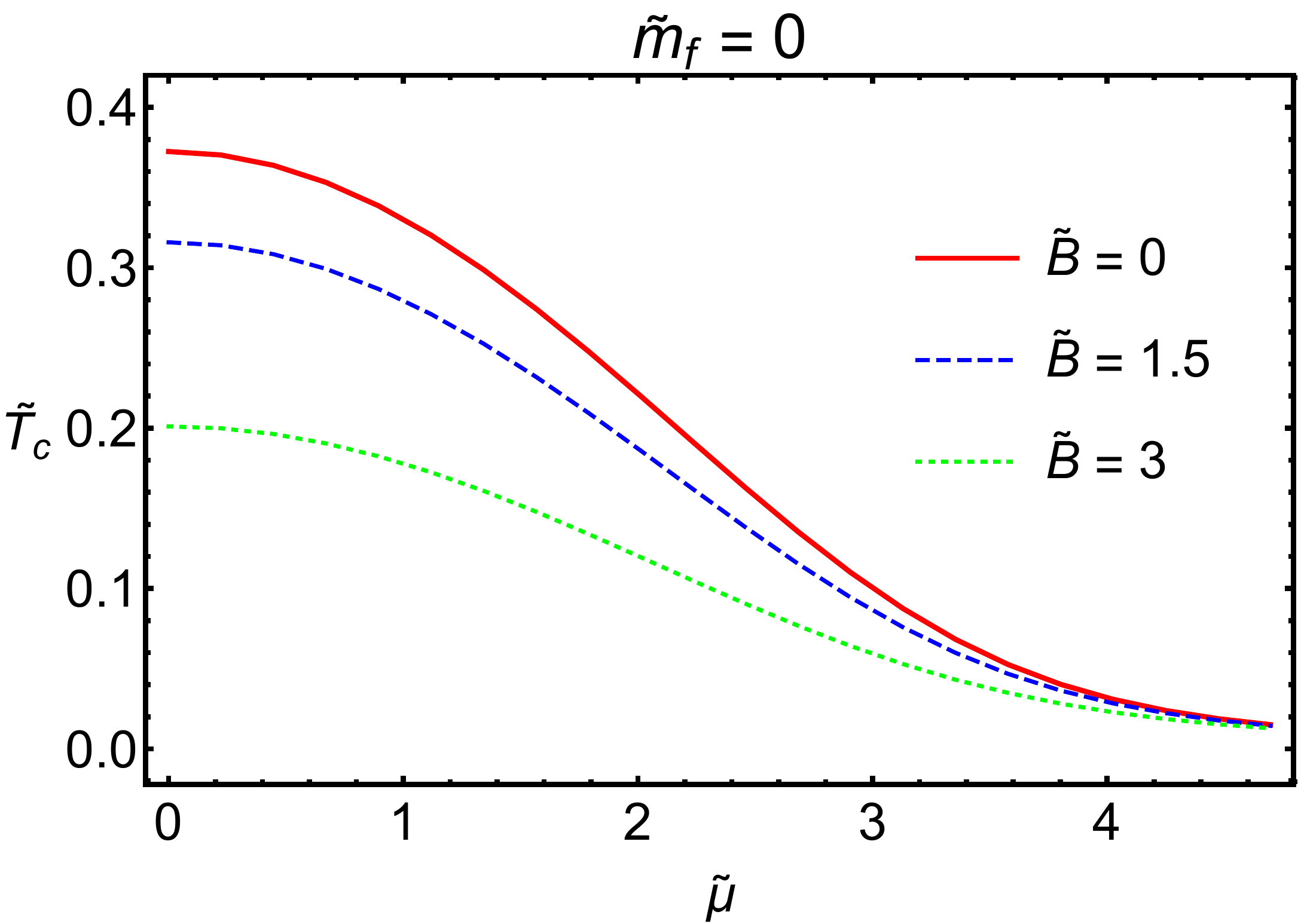}
	\hfill
	\includegraphics[scale = 0.34]{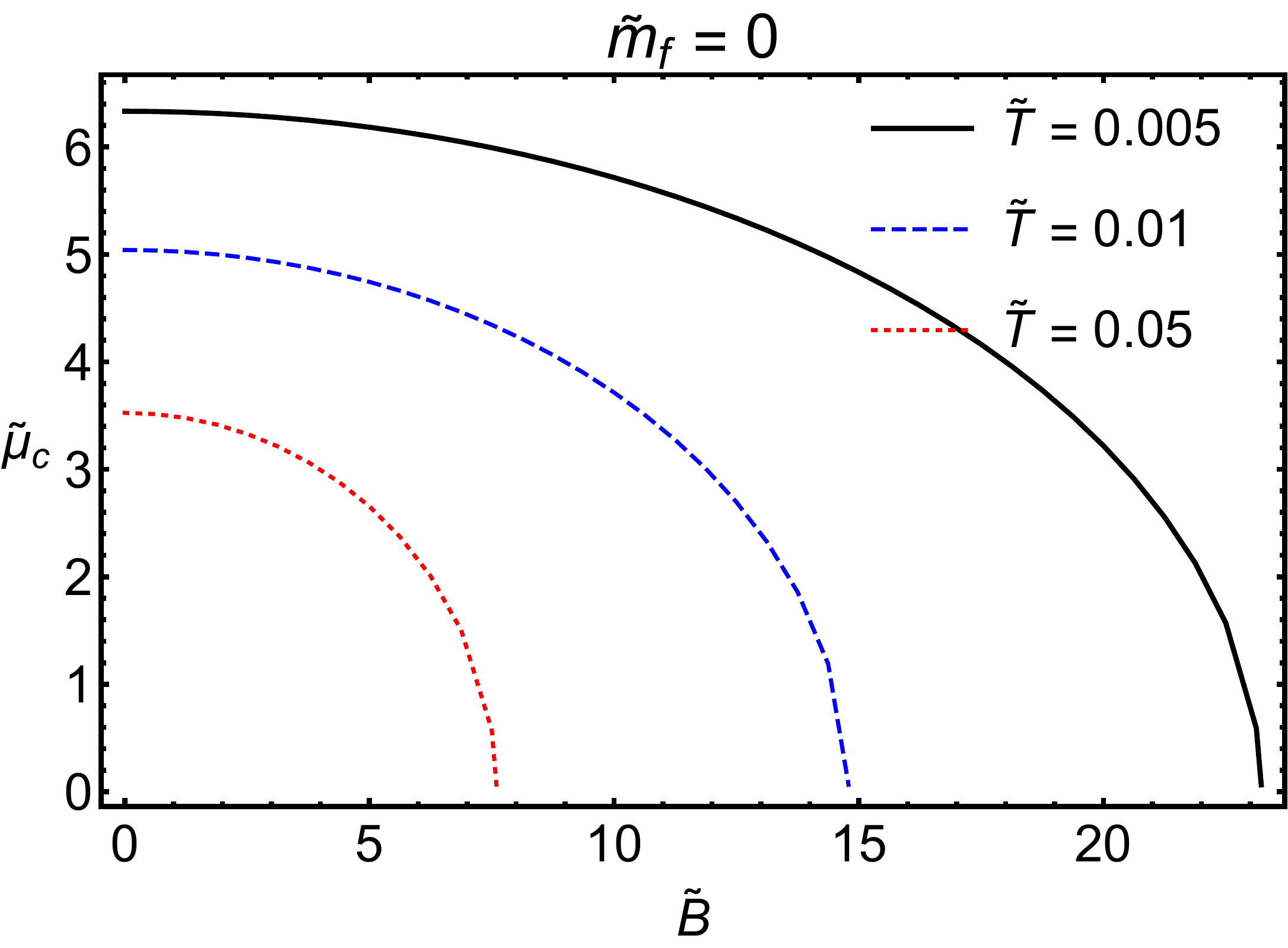}
	\caption{{\sl Left panel}: critical temperature $\tilde T_{c} $ versus the chemical potential $\tilde \mu$ in the chiral limit for different values of the magnetic field. The critical temperature is defined as the temperature where the chiral condensate vanishes for fixed $\mu$ and $B$. {\sl Right panel}: critical chemical potential $\tilde \mu_{c} $ versus $\tilde B$ in the chiral limit for different values of temperature. As for the critical temperature, the critical chemical potential is defined as the chemical potential where the chiral condensate vanishes for fixed $T$ and $B$. All quantities are in units of $\sqrt{\phi_{\infty}}$, in both panels.}
	\label{fig:svsTc}
\end{figure}

Finally, in Figure \ref{fig:svsTc} is presented the critical temperature $\tilde T_{c} $ as a function of the chemical potential $\tilde \mu$ in the chiral limit for different values of magnetic field ({\sl Left panel}) and the critical chemical potential $\tilde \mu_{c} $ versus $\tilde B$ in the chiral limit for different values of the temperature ({\sl Right panel}). These critical quantities ($T_c$,\,$\mu_c$) are defined as follows. The critical temperature $T_c$ is defined as the temperature where the chiral condensate vanishes for fixed $\mu$ and $B$. Analogously, the critical chemical potential $\mu_c$ is defined as the chemical potential where the chiral condensate vanishes for fixed $T$ and $B$.
These findings give additional support to the fact that our holographic model captures the IMC effect at zero and finite densities as well as a decrease on the chiral condensate with increasing chemical potential with or without magnetic fields. We also find that the two effects related to the presence of chemical potential and magnetic fields on the chiral condensate sum up decreasing the chiral condensate even more.

\section{Conclusions}\label{conclusions}

In this work we have described holographically finite density effects on the spontaneous chiral symmetry breaking and chiral phase transition of a system in $ 2+1 $ dimensions in the presence of magnetic fields. We observe inverse magnetic catalysis (IMC), which is the reduction of the chiral condensate with an increasing magnetic field, at zero or at finite density. We also observe a decreasing of the chiral condensate with increasing chemical potential, with or without magnetic fields. Furthermore, the reduction of the chiral condensate is even more pronounced when one takes both finite densities and magnetic fields simultaneously, as shown in Figures \ref{fig:svsTa} and \ref{fig:svsTb}. 
Moreover, we have also find that the critical temperature $\tilde T_c$ diminishes with increasing chemical potential and that the critical chemical potential $\tilde \mu_c$ decreases with increasing magnetic field, as pictured in Figure \ref{fig:svsTc}. These results are in good agreement with other higher-dimensional holographic studies, such as the one presented in Refs. 
\cite{Gursoy:2017wzz, Ballon-Bayona:2017dvv, Gursoy:2018ydr}. 

As a possible extension to our holographic model, it would be interesting to include a Dirac-Born-Infeld (DBI) action in which the magnetic field and the tachyon are coupled. In this setup one might reproduce magnetic catalysis (MC) in our holographic model along with the standard inverse magnetic catalysis (IMC) which comes from the contribution of the magnetic field introduced via the metric. A possible clue in this direction is given by the recent higher-dimensional holographic analysis presented in Ref. \cite{Ballon-Bayona:2020xtf} at zero density where there is MC, as would be expected from QCD. 

Another possible extension for our work is to consider 
the confined phase for low temperatures besides the deconfined phase for high temperatures, separated by a Hawking-Page phase transition. In this case we could study the deconfinement phase transition together with the chiral phase transition which are expected to happen approximately at the same temperature in QCD.

\section*{Acknowledgments}
The authors thank Alfonso Ballon Bayona and Luis Mamani for useful conversations. DMR is supported by Conselho Nacional de Desenvolvimento Científico e Tecnológico (CNPq) under grant No. 152447/2019-9. D.L. is supported by the National Natural Science Foundation of China (11805084), the PhD Start-up Fund of Natural Science Foundation of Guangdong Province (2018030310457) and Guangdong Pearl River Talents Plan (2017GC010480). H.B.-F. is partially supported by Coordenação de Aperfeiçoamento de Pessoal de Nível Superior (CAPES), and Conselho Nacional de Desenvolvimento Científico e Tecnológico (CNPq) under Grant No. 311079/2019-9.

\end{document}